\documentclass[10pt]{article}
\usepackage{graphicx,amssymb,epsfig}
\def\issue(#1,#2,#3){{\bf #1}, #2 (#3)} 

\def\opcit(#1){ {\em op. cit.}, #1}

\def\APP(#1,#2,#3){Acta Phys.\ Polon.\ \issue(#1,#2,#3)}
\def\ARNPS(#1,#2,#3){Ann.\ Rev.\ Nucl.\ Part.\ Sci.\ \issue(#1,#2,#3)}
\def\CPC(#1,#2,#3){Comp.\ Phys.\ Comm.\ \issue(#1,#2,#3)}
\def\CIP(#1,#2,#3){Comput.\ Phys.\ \issue(#1,#2,#3)}
\def\EPJC(#1,#2,#3){Eur.\ Phys.\ J.\ C\ \issue(#1,#2,#3)}
\def\EPJD(#1,#2,#3){Eur.\ Phys.\ J. Direct\ C\ \issue(#1,#2,#3)}
\def\IEEETNS(#1,#2,#3){IEEE Trans.\ Nucl.\ Sci.\ \issue(#1,#2,#3)}
\def\IJMP(#1,#2,#3){Int.\ J.\ Mod.\ Phys. \issue(#1,#2,#3)}
\def\JHEP(#1,#2,#3){J.\ High Energy Physics \issue(#1,#2,#3)}
\def\MPL(#1,#2,#3){Mod.\ Phys.\ Lett.\ \issue(#1,#2,#3)}
\def\NP(#1,#2,#3){Nucl.\ Phys.\ \issue(#1,#2,#3)}
\def\NIM(#1,#2,#3){Nucl.\ Instrum.\ Meth.\ \issue(#1,#2,#3)}
\def\PL(#1,#2,#3){Phys.\ Lett.\ \issue(#1,#2,#3)}
\def\PR(#1,#2,#3){Phys.\ Rev.\ \issue(#1,#2,#3)}
\def\PRL(#1,#2,#3){Phys.\ Rev.\ Lett.\ \issue(#1,#2,#3)}
\def\SJNP(#1,#2,#3){Sov.\ J. Nucl.\ Phys.\ \issue(#1,#2,#3)}
\def\ZPC(#1,#2,#3){Zeit.\ Phys.\ C \issue(#1,#2,#3)}
\def\JPG(#1,#2,#3){J.\ Phys.\ G \issue(#1,#2,#3)}

\def\bra {\langle}
\def\ket {\rangle}
\def\bfl{\begin{flushleft}}
\def\efl{\end{flushleft}}

\def\delms {\Delta {M_s}}

\def\bar {\overline}
\def\bbbar {B_{q}^0-\overline{B_q}{}^0}
\def\kkbar {K^0-\bar{K}{}^0}
\def\ddbar {D^0-\bar{D}{}^0}
\def\mmbar {M^0-\bar{M}{}^0}
\def\bsbsbar {B_s-\bar{B_s}}
\def\bdbdbar {B_d-\bar{B_d}}

\def\be {\begin{equation}}
\def\ee {\end{equation}}
\def\bea {\begin{eqnarray}}
\def\eea {\end{eqnarray}}

\def\bc {\begin{center}}
\def\ec {\end{center}}

\def\delms {\Delta {M_{B_s}}}
\def\delmbd {\Delta {M_{B_d}}}
\def\delmk {\Delta {M_{K}}}

\begin{document}

\begin{flushright}
IMSC-PHYSICS/08-2009\\
CU-PHYSICS/2-2010\\
\end{flushright}

\begin{center}
{\Large \bf Constraining Scalar Leptoquarks from the K and B Sectors}\\
\vspace*{1cm}
\renewcommand{\thefootnote}{\fnsymbol{footnote}}
{\large {\sf Jyoti Prasad Saha ${}^1$}
\footnote{E-mail: jyotiprasadsaha@gmail.com},
{\sf Basudha Misra ${}^2$}
\footnote{E-mail: basudha@imsc.res.in},
and {\sf Anirban Kundu ${}^3$}
\footnote{E-mail: akphy@caluniv.ac.in}
} \\
\vspace{10pt}
{\small
${}^{1)}$ {\em Department of Physics, Serampore College, Serampore 712201,
India }\\
${}^{2)}$ {\em Institute of Mathematical Sciences, Chennai 600113, India\\
and\\
Centre for High Energy Physics, Indian Institute of Science, Bangalore 560012,
India}\\
${}^{3)}$ {\em Department of Physics, University of Calcutta, 92 A.P.C.
Road, Kolkata 700009, India}}
\normalsize
\end{center}

\date{\today}

\begin{abstract}
Upper bounds at the weak scale are obtained for all $\lambda_{ij}\lambda_{im}$ 
type product couplings of the scalar leptoquark model which may affect
$\kkbar$, $\bdbdbar$ and $\bsbsbar$
mixing, as well as leptonic and semileptonic K and B decays.
Constraints are obtained for both real and imaginary parts of the couplings. We also
discuss the role of leptoquarks in explaining the anomalously large CP-violating
phase in $\bsbsbar$ mixing. 
\end{abstract}

\bfl
\vspace*{0.05in}
\efl


\section{Introduction}

The Standard Model (SM), in all probability, is just an effective theory valid
up to a scale which is much below the Planck scale, and
hopefully in the range of a few hundreds of GeV, so that the physics beyond SM
can be explored at the LHC. Direct production of new particles will definitely
signal new physics (NP); while it is an interesting problem to find out what
type of NP is there (commonly known as the `inverse problem'), 
it is also well-known that indirect data
from low-energy experiments will help to pin down the exact structure of NP,
including its flavour sector. The low-energy data, in particular the data
coming from the B factories as well as from CDF, D\O, LHCb
(and also from the general
purpose ATLAS and CMS experiments) are going to play
a crucial role in that.
There are already some interesting hints; just to name a few \cite{hfag}: 
(i) the large mixing phase in $\bsbsbar$ mixing; 
(ii) the fraction of longitudinally 
polarised final states in channels like $B\to \phi K^\ast$ and $B\to \rho K^\ast$;
(iii) the anomalous direct
CP-asymmetries in $B\to \pi K$ decays;
(iv) the discrepancy in the extracted value of $\sin(2\beta)$ from
$B_d\to J/\psi K_S$ and $B_d\to \phi K_S$; (v) the larger branching fraction
of $B^+\to\tau^+\nu$ compared to the SM expectation; and 
(vi) the discrepancy in the extracted values of $V_{ub}$ from inclusive and
exclusive modes. While none of them are conclusive 
proof of any NP, there is a serious tension with the SM when all the data are
taken together. If all, or most, of them survive the test of time and attain 
more significance, this will indicate a new physics whose flavour sector
is definitely of the non-minimal flavour violating (NMFV) type.

In this paper, as an example of an NMFV new physics, 
we focus upon the model of scalar leptoquarks (LQ). 
In general, as in any NMFV model, we expect possibly large
deviations from the SM in the flavour sector observables.
LQs that violate both baryon number B and lepton number L must be massive
at the level of $\sim 10^{15}$ GeV to avoid proton decay, and are of no interest to us.
(There are exceptions; one can construct models where LQs violate both
B and L and yet do not mediate proton decay. These LQs can be light.
For example, see \cite{dorsner1}.)
On the other hand, LQs conserving either B or L or both can be light, ${\cal O}(100$
GeV), and we discuss the phenomenology of only those models that conserve both
B and L; one can find extensive discussions on these models in  \cite{sacha,leurer1,leurer2}. 
Vector LQs, as well as some gauge-nonsinglet scalar ones, 
couple to neutrinos, and their couplings should be very tightly constrained from
neutrino mass and mixing data. 

Another phenomenological motivation for a LQ model is that
this is one of the very few models (R-parity violating supersymmetry is another)
where the neutral meson mixing diagram gets a new contribution to the absorptive
part. This, for example, may lead to an enhancement in the width difference
$\Delta\Gamma$ in the $B_s$ system \cite{dighe}, 
contrary to what happens in more popular
NP models that can only decrease $\Delta\Gamma$ \cite{grossman}. The NP also
changes the CP-violating phase in $B_s\to
J/\psi \phi$ and hence can help reducing
the tension \cite{cdf-bs} 
of SM expectation and the Tevatron data on the CP-violating phase and
width difference for $B_s$. 

All flavour-changing observables constrain the product of at least two different
LQ couplings, one linked with the parent flavour and another with the
daughter flavour. The product couplings may be complex and it is generally 
impossible to absorb the phase just by a simple redefinition of the LQ field.
We use the data from $\kkbar$, $\bsbsbar$ and  $\bdbdbar$ mixing to constrain the relevant product
couplings, generically denoted as $\lambda\lambda$. 
For the $B$ system, we use the data on $\Delta M_{d,s}$ and
the mixing phase $\sin(2\beta_{d,s})$, and for the $K$ system, we use 
$\Delta M_K$ and $\varepsilon_K$. We do not discuss other CP
violating parameters like $\varepsilon'/\varepsilon$, since that has large
theoretical uncertainties.  
We also discuss the correlated leptonic and semileptonic decays, i.e., the
decays mediated by the same LQ couplings. 
While decays to most of the semileptonic
channels have been observed, 
the clean leptonic channels only have an upper bound for almost all the cases, except the 
already observed leptonic decays
$K_L\to e^+e^-, \mu^+\mu^-$. Note that the final
state must have leptons and hence the bounds are more robust compared to those coming from models with
only hadrons in the final state.

A similar exercise have also been 
undertaken in \cite{sacha,campbelle6,dimopoulos,study1,study2}. We update these bounds with new data from the B factories and other collider experiments. 
In particular, in the subsequent sections, all the previous
bounds that we quote have been taken from \cite{sacha}.
The $\ddbar$ system has not been considered due to the
large theoretical uncertainties and dominance of long-distance contributions. We refer the reader to \cite{pakvasa} for a discussion on the bounds coming from $\ddbar$ mixing. Leptonic and semileptonic D and $D_s$ decays have been used to put constraints on LQs that couple to the up-type quarks.
In particular, LQ contribution might be interesting to explain the $D_s$ leptonic decay anomaly  
\cite{dorsner1,bogdan,fajfer}. The couplings that we constrain are generically of the type $\lambda_{ij}
\lambda_{ik}^\ast$, where the $k$-th quark flavour changes to the $j$-th, but there is no flavour
change in the lepton sector. One can, in principle, consider flavour changes in the lepton sector
too; for an analysis of that type of processes, see \cite{smirnov}. However, if one has a $\nu\bar\nu$
pair in the final state, as in $K_L\to \pi^0\nu\bar\nu$, there is a chance that lepton flavour is
also violated.

The couplings, which are in general complex, may be constrained from a 
combined study of CP-conserving and CP-violating observables. For neutral mesons,
these mean $\Delta M$ as well as $\epsilon_K$ and $\sin(2\beta_{d,s})$.
However, for most of the cases, the leptonic and semileptonic decay channels
provide the better bound. 
The analysis
has been done keeping both the SM and LQ contributions, which keeps the possibility of a destructive
intereference, and hence larger possible values of the LQ amplitudes, open. 

In Section 2 we briefly state the relevant formulae necessary for the analysis. Section 3 deals with the numerical inputs.
In Section 4, we take up the analysis, first of the neutral meson mixing, and then the correlated leptonic
and semileptonic decays. We
conclude and summarize in Section 5.


\section{Relevant Expressions} 
\subsection{Neutral Meson Mixing}

For the neutral meson system generically denoted by $M^0$ and $\overline{M}{}^0$, 
the mass difference between the two mass eigenstates $\Delta M$ is given by
\begin{eqnarray}
\Delta M&=&2{\rm Re}\left[ (M_{12}-{i\over 2}\Gamma_{12}) 
(M^*_{12}-{i\over 2}\Gamma^*_{12})\right]^{1/2}\,,\nonumber\\
\Delta \Gamma&=&-4{\rm Im}\left[ (M_{12}-{i\over 2}\Gamma_{12}) 
(M^*_{12}-{i\over 2}\Gamma^*_{12})\right]^{1/2}\,.
\end{eqnarray}
For the B system, $|M_{12}|\gg |\Gamma_{12}|$ and  $\Delta M = 2|M_{12}|$.
For the K system, if the decay is dominantly to the $I=0$, ${\rm Im}\Gamma_{12}$
can be neglected and one can write
\begin{equation}
|\varepsilon_K| = {1\over 2\sqrt{2}} {{\rm Im}M_{12}\over {\rm Re} M_{12}}
 = {1\over \sqrt{2}} {{\rm Im} M_{12}\over \Delta M_K}\,.
    \label{epsa-k}
\end{equation}

Let the SM amplitude be
\begin{equation}
M_{12}^{SM} = |M_{12}^{SM}|\exp(-2i\theta_{SM})\,,
\end{equation}
where $\theta_{SM}= \beta_d$ for the ${\bdbdbar}$ system and 
approximately zero
for $\kkbar$ and $\bsbsbar$ systems. 

If we have $n$ number of new physics (NP)
amplitudes with weak phases $\theta_n$, one can write
\begin{equation}
M_{12} = |M_{12}^{SM}|\exp(-2i\theta_{SM}) + \sum_{i=1}^n
|M_{12}^i|\exp(-2i\theta_i)\,.
\end{equation}
This immediately gives the effective mixing phase $\theta_{\it {eff}}$ as
\begin{equation}
\theta_{\it {eff}} = {1\over 2}\arctan {|M_{12}^{SM}|\sin(2\theta_{SM}) + 
\sum_i|M_{12}^i|\sin(2\theta_i)
\over |M_{12}^{SM}|\cos(2\theta_{SM}) + \sum_i|M_{12}^i|\cos(2\theta_i)},
\end{equation}
and the mass difference between mass eigenstates as
\begin{eqnarray}
\Delta M & = & 2 [ |M_{12}^{SM}|^2 + \sum_i|M_{12}^i|^2
 + 2|M_{12}^{SM}|\sum_i |M_{12}^i|\cos 2(\theta_{SM}-\theta_i)\nonumber \\
& + & 2 \sum_i \sum_{j>i} |M_{12}^j||M_{12}^i|\cos 2(\theta_j-\theta_i)]^{1/2}\,.
\end{eqnarray}

For the $\kkbar$ system, the dominant part of the short-distance SM amplitude is
\be
M_{12}^{SM}\equiv {\bra \bar{K^0}|H_{eff}|K^0\ket\over  2m_K}
\approx {G_F^2\over 6\pi^2}(V_{cd}V_{cs}^*)^2
\eta_K m_K f_K^2 B_K m_W^2 S_0(x_c)\,,
    \label{k-sm}
\ee
where $x_j = m_j^2/m_W^2$, $f_K$ is the $K$ meson decay constant, and
 $\eta_K$ (also called $\eta_{cc}$ in the literature)
and $B_K$ parametrize the short- and the long-distance QCD
corrections respectively. The top-quark loop dependent part, which is tiny due to
the CKM suppression, but responsible
for CP violation, has been neglected. 
The function $S_0$ is given by
\begin{equation}
S_0(x) = {4x-11x^2+x^3\over 4(1-x)^2} - {3x^3\ln x \over 2(1-x)^3}\,.
\end{equation}
For the $\bbbar$ system ($q=d$ for $\bdbdbar$ and $q=s$ for $\bsbsbar$), 
we have an analogous equation, dominated by the top quark loop:
\be
M_{12}^{SM}\equiv {\bra \bar{B_q^0}|H_{eff}|B_q^0\ket\over  2m_{B_q}}
= {G_F^2\over 6 \pi^2}(V_{tq}V_{tb}^*)^2
\eta_{B_q} m_{B_q} f_{B_q}^2 B_{B_q} m_W^2 S_0(x_t)\,.
    \label{b-sm}
\ee
\subsection{Leptonic and Semileptonic Decays}

For almost all the cases, the SM leptonic decay widths for neutral mesons
are way too small to be 
taken into account, and we can safely saturate the present bound with the
NP amplitude alone, except for the $K_L$ sector. For example, the branching ratio
of $B_s\to\mu^+\mu^-$ is about $3.4\times 10^{-9}$ and that of $B_d\to \mu^+
\mu^-$ is about $1.0\times 10^{-10}$ in the SM, while the experimental limits are
at the ballpark of 4-6$\times 10^{-8}$.
Another exception is the $B^-\to l^-\bar\nu$ decay, which proceeds through
the annihilation channel in the SM:
\be
{\rm Br}(B^-\to l^-\bar\nu) = \frac{1}{8\pi} G_F^2 m_B m_l^2 f_B^2 |V_{ub}|^2
\tau_B \left( 1 - \frac{m_l^2}{m_B^2}\right)^2\,,
\label{b-lepdecay}
\ee
where $\tau_B$ is the lifetime of the B meson.  

For the semileptonic decays, we use the following standard
convention \cite{F0}, given for the $B\to K^{(\ast)}\ell^+\ell^-$ transition:
\bea
\bra K(p_2) | \bar{b}\gamma_\mu s|B(p_1)\ket &=&
P_\mu F_1(q^2) +q_{\mu} \frac{m_B^2-m_K^2}{q^2}
\left(F_0(q^2)-F_1(q^2)\right),
\nonumber\\
\langle K^\ast(p_2,\epsilon ) | \bar{b} \gamma_\mu(1\mp \gamma_5) s \ |B(p_1)
         \rangle &=&
         \mp iq_\mu \frac{2m_{K^\ast}}{q^2} \epsilon^\ast.q \left[A_3(q^2)-A_0(q^2)\right]\nonumber\\
         && \pm i\epsilon_\mu^\ast (m_{B}+m_{K^\ast}) A_1(q^2) \mp \frac{i}{m_{B}+m_{K^\ast}} 
         P_\mu (\epsilon^\ast.q) A_2(q^2) \nonumber\\
         && -\varepsilon_{\mu\nu\alpha\beta} \epsilon^{\ast\nu} p_2^\alpha q^\beta \frac{2V(q^2)}
         { m_{B}+m_{K^\ast} }\,,         
\label{Bphi} 
\eea
 where 
 $P=p_1+p_2$, and $q=p_1-p_2$. The pole dominance ensures that $A_3(0)=A_0(0)$, and 
 $A_3(q^2)$ can be expressed in terms of $A_1$ and $A_2$.

\subsection{Leptoquarks}


Leptoquarks are colour-triplet bosons
that can couple to a quark and a lepton at the same time, and can occur in a number
of Grand Unified Theories (GUTs) \cite{paticonstrain}, composite models 
\cite{comp1}, and superstring-inspired $E_6$ models \cite{campbelle6}.
In fact, the R-parity violating squarks of supersymmetry, as far as
their couplings with fermions go, are nothing but LQs.  Model-independent
constraints on their properties are available \cite{sacha,leurer1,leurer2}.

We focus on the scalar LQ model, which conserves both B and L.
The relevant part of the Lagrangian \cite{sacha} can be written as 
\begin{equation}
\label{eq:lepto-lag}
\begin{array}{rcl}
{\cal L}_S & = & \left\{ \right. ( \lambda_{LS_0} \bar{q}^c_L i\sigma_2 l_L
+ \lambda_{RS_0} \bar{u}^c_R e_R) S_0^{\dagger} + \lambda_{R \tilde{S_0}}
\bar{d}_R^c e_R \tilde{S}_0^{\dagger}  + (\lambda_{LS_{\frac12}}
\bar{u}_R l_L +
  \\ & &  \lambda_{RS_{\frac12}}\bar{q}_L  i\sigma_2  e_R) S_{\frac12}^{\dagger} +
 \lambda_{L \tilde{S}_{\frac12}}
\bar{d}_R l_L \tilde{S}_{\frac12}^{\dagger}
 + \lambda_{LS_{1}} \bar{q}^c_L i\sigma_2 {\sigma^a}
l_L \cdot {S^a_1}^{\dagger}  \left. \right\} + h.c.
\end{array}
\end{equation}
where $(S_0,\tilde{S}_0)$,  $(S_{\frac12},\tilde{S}_{\frac12})$, and
 $S^a_1$ ($a=1,2,3$) represent the $SU(2)$ singlet, doublet,
and triplet LQs respectively. 
$\lambda^{ij}$ is the coupling strength of
a leptoquark to an $i$-th generation lepton and a $j$-th generation quark, which is in general complex. 
$\sigma$'s are the Pauli spin matrices. Note that all the four
terms that couple a neutrino with a LQ can have potential constraints on neutrino
mass and mixing. For example, $\tilde{S}_{\frac12}$ can generate the observed neutrino
mixing pattern through a type-II seesaw mechanism \cite{dorsner2,perez}. However, the 
constraints also depend on the vacuum expectation value of a higher-representation
scalar field. That is why we show the non-neutrino constraints for these couplings too,
keeping in mind that the neutrino constraints may be stronger. 

In this work, we focus only on those processes that involve down-type quarks. Thus, 
there is no way to constrain $\lambda_{RS_0}$ and $\lambda_{LS_{\frac12}}$ from these
processes. In fact, these two sets of coupling can be constrained from processes like
$D^0$-$\overline{D}{}^0$ mixing and $\ell_i\to \ell_j+\gamma$. The latter can be constrained from 
neutrino mixing too, but as we have just mentioned, the limits would depend on other
model parameters. 
We constrain only five types of scalar LQ couplings here: $\lambda_{LS_0}$, $\lambda_{R
\tilde{S_0} }$, $\lambda_{RS_{\frac12}}$, $\lambda_{L\tilde{S}_{\frac12}}$, and $\lambda_{LS_1}$.

While we have not explicitly shown the generation indices in eq. (\ref{eq:lepto-lag}), it is assumed that the LQs can couple 
with fermions from two di?erent generations. There is another approach which we should mention here. In 
this approach \cite{leurer1,leurer2}, one takes the LQ coupling for a single generation, say the third, so that only the third 
generation fermions are a?ected. However, these couplings are taken to be in the weak basis, and when one 
rotates to the mass basis of the quarks, the cross-generational couplings are generated with a mechanism similar 
to that of the Cabibbo-Kobayashi-Maskawa (CKM) mixing. This controls the relative magnitudes, as well as 
the phases, of the LQ couplings, and given a texture in the weak basis, all the couplings in the mass basis are 
present, with predicted magnitudes and phases (the LQ coupling phase also comes from the CKM phase). 

While one gets comparable bounds to that of \cite{sacha} in this scheme as well, one should note that: \\
(i) The mixing matrix for the down-type quarks is not known. What one knows is the misalignment between the 
$u_L$ and the $d_L$ bases. Thus, one is forced to consider only the charged-current processes where the misalignment 
(and not the individual rotation matrices) matters. \\
(ii) The CKM scheme does not say anything about the rotation matrices for the right-chiral quark sector. 
Whatever one uses there is at best an assumption. A similar analysis has been done for R-parity violating 
supersymmetric models too \cite{dreiner}. 

Thus, we will assume that whatever couplings are nonzero, are so in the physical basis of the quark Þelds, and 
the phase is arbitrary and not a function of the CKM parameters.

\subsection{Direct Production Limits}

The direct production limits depend on the LQ model, as well as the SM fermions these LQs can 
couple to. 
The best limits are as follows \cite{pdg}: $m_{LQ} > 256, 316, 229$ GeV for 1st, 2nd, and 3rd
generation LQs respectively when they are pair produced, and $m_{LQ} > 298, 73$ GeV 
for 1st and 2nd generation LQs when there is single production. These are the absolute
lower bounds at 95\% CL, but the constraints are tighter if the LQ coupling to the SM
fermions, denoted here by $\lambda$, is large. For example, if $\lambda$ is of the
order of the electromagnetic coupling, the first generation LQs were found to have a 
limit of 275-325 GeV by both the HERA experiments \cite{h1}. For even larger couplings
($\lambda = 0.2$-$0.5$ and above) the LEP experiments exclude a much wider mass
range \cite{opal}, 
but such strong couplings are already almost ruled out if LQ signals are to be
observed at the LHC. 

The present-day limits for pair-produced LQs are due to the Tevatron experiments. For
a summary, we refer the reader to \cite{recent-exp}. The first generation LQs are searched
in $2e+2j$ or $1e+2j+$MET channel; the second generation ones are in $2\mu+2j$ or
$1\mu+2j+$MET channel; and the third generation ones are in $2\tau+2b$ or $2b+$MET
channel. We refer to the original papers \cite{cdf-d0} for the details of the analysis. 

The production of LQ states, either single, associated with a lepton (from 
$qg \to LQ+\ell$), or in pair, from $q\bar q,gg\to LQ+\overline{LQ}$, has been studied
in detail; for example, the reader may look at \cite{perez,blumlein,mitsou,belyaev}. 
At $\sqrt{s}=14$ TeV, the cross-section of pair production of scalar leptoquarks is about
1 pb \cite{belyaev}. While this goes down significantly for the initial LHC run of $\sqrt{s}=7$ 
TeV, one expects to see LQ signals upto $m_{LQ}=500$ GeV even with 5 fb$^{-1}$ of
luminosity. The cross-section for single production depends on the value of
$\lambda$. For $\lambda=\sqrt{4\pi\alpha}$, the electric charge, the cross-section 
for LQ plus charged lepton production is about 100 fb for $m_{LQ}=500$ GeV. The
cross-section is proportional to $\lambda^2$, so we expect events even with 
$\sqrt{s}=7$ TeV and $\lambda \approx 0.05$. Obviously, for smaller values of 
$\lambda$, pair production will be more favoured, and we expect the preliminary run
of LHC to establish a limit of the order of 500 GeV. 

In this analysis,
we will use a somewhat conservative reference mass value of 300 GeV for every LQ state,
independent of the quantum numbers and generation. The bounds on the product couplings
scale as $m_{LQ}^2$, so the bounds that we show should be multiplied by 
$(300/m_{LQ})^2$.

\subsection{Constraints from Meson Mixing}
\label{subsec:lqmm}

\begin{figure*}
\begin{center}
\centerline{\hspace*{3em}
\epsfxsize=16cm\epsfysize=4.5cm
                     \epsfbox{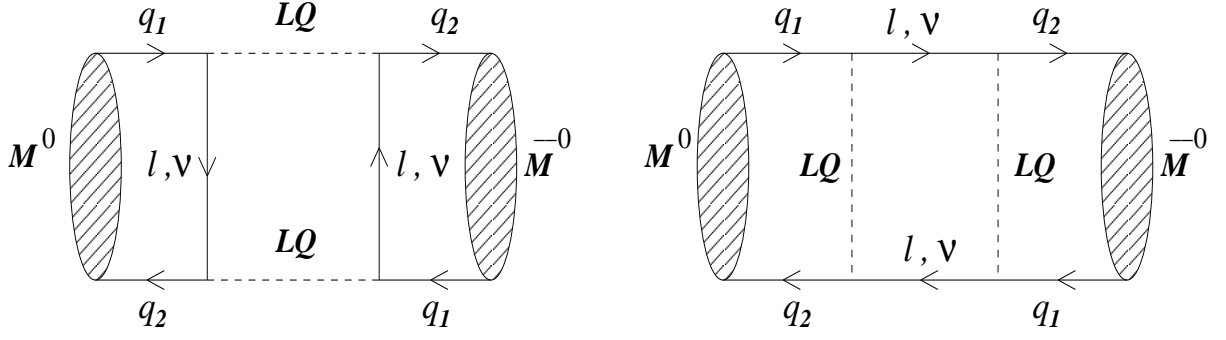}
}
\end{center}
\hspace*{-3cm}
  \caption{\em Leptoquark contributions to $\mmbar$ mixing.}
 \label{fig:boxdiag}
\end{figure*}

Consider the neutral meson $M^0\equiv q_j\bar{q_k}$. The oscillation can have a new 
LQ mediated amplitude, with $i$-type leptons and some scalar LQs in the box, as
shown in Fig.\ (\ref{fig:boxdiag}). The amplitude is proportional to 
$(\lambda^\ast_{ik}\lambda_{ij})^2$. We consider, as in the standard practice,
a hierarchical coupling scheme, so that we may consider only two LQ couplings
to be nonzero at the most. Also, we consider any one type of LQ to be present
at the same time. This keeps the discussion simple and the numerical results
easily tractable; however, this may not be the case where we have some high-energy
texture of the couplings and there can be a number of nonzero couplings at the
weak scale.

For the LQ box, one must consider the same type of lepton flowing inside
the box if we wish to restrict the number of LQ couplings to 2. The effective
Hamiltonian contains the operator $\tilde{O}_1$, defined as
\be
\tilde{O}_1 =   \left[\bar{b}\gamma^\mu P_R d\right]_1  \left[\bar{b}\gamma_\mu P_R d\right]_1 \,,
\ee
(where the subscript 1 indicates the SU(3)$_c$ singlet nature of the current), and is given by
\be
{\cal H}_{LQ} = {({\lambda^\ast_{ik}}\lambda_{iq})^2\over 128 \pi^2}
\left[\frac{c_1}{m_{LQ}^2} \left\{ I\left({m_l}^2\over 
m_{LQ}^2\right)\right\}+ 
\frac{c_2}{m_{LQ}^2} \right]\tilde{O_1}\,,
\ee
where $c_1=1$, $c_2=0$ for $S_0,\tilde{S}_0$ and $S_{\frac12}$, $c_1=c_2=1$ for $\tilde{S}_\frac12$, and
$c_1=4$, $c_2=1$ for $S_1$. Therefore, if we are allowed to neglect the SM, the limits on the 
product couplings for 
$(\lambda_{LS_0}, \lambda_{R\tilde{S_0} }, \lambda_{RS_{1/2}})$, $\lambda_{L\tilde{S}_{1/2}}$, and $\lambda_{LS_1}$
should be at the ratio of $1:\frac{1}{\sqrt{2}}:\frac{1}{\sqrt{5}}$. 
The operator $\tilde{O}_1$ is multiplicatively renormalized
and the LQ couplings are those obtained at the weak scale. The function $I(x)$, defined as
\begin{equation}
I(x) = {1-x^2+2x\log x\over (1-x)^3}\,,
\end{equation}
is always very close to $I(0)=1$; note that we have taken all LQs to
be degenerate at 300 GeV.


\subsection{Constraints from Leptonic and Semileptonic Decays}
\label{sec:2bodydecay}

\begin{table}[htbp]
\begin{center}
\begin{tabular}{||c|c|c||}
\hline {Interaction} & {4-fermion vertex} &
{Fierz-transformed vertex}\\ 
\hline 

($\lambda_{LS_0}\bar{q}^c_L i\sigma_2 l_L + \lambda_{RS_0}
\bar{u}^c_R e_R) S_0^{\dagger}$&
$G (\overline{d}_{L}^c
\nu_{L}) (\overline{\nu}_{L}d_{L}^c)$ &
$\frac12 G 
(\overline{d}_{L}^c\gamma^{\mu}d_{L}^c)
(\overline{\nu}_{L}\gamma_{\mu}\nu_{L})$\\

\hline

 $\lambda_{R \tilde{S_o}}
\bar{d}_R^c e_R \tilde{S}_0^{\dagger} $ &
$G
(\overline{d}_{R}^c e_{R}) (\overline{e}_{R}d_{R}^c)$ &
$\frac12 G
(\overline{d}_{R}^c\gamma^{\mu}d_{R}^c)
(\overline{e}_{R}\gamma_{\mu}e_{R})$\\

\hline 

$(\lambda_{LS_{1/2}}
\bar{u}_R l_L +\lambda_{RS_{1/2}}\bar{q}_L  i\sigma_2  e_R) S_{1/2}^{\dagger}$ &
$G  (\overline{d}_{L}e_{R})
(\overline{e}_{R}d_{L})$ & $\frac12 G
 (\overline{d}_{L}\gamma_{\mu}d_{L})
(\overline{e}_{R}\gamma_{\mu}e_{R})$\\

\hline 

$\lambda_{L \tilde{S}_{1/2}} \bar{d}_R l_L
\tilde{S}_{1/2}^{\dagger}$ &
 $G
(\overline{d}_{R}\nu_{L}) (\overline{\nu}_{L}d_{R})$ &
$\frac12 G
(\overline{d}_{R}\gamma^{\mu}d_{R})
(\overline{\nu}_{L}\gamma_{\mu}\nu_{L})$\\

 & $G
  (\overline{d}_{R}e_{L})
(\overline{e}_{L}d_{R})$ & $\frac12 G
(\overline{d}_{R}\gamma^{\mu}d_{R})
(\overline{e}_{L}\gamma_{\mu}e_{L})$\\

\hline 

$\lambda_{LS_{1}}
\bar{q}^c_L i\sigma_2 \vec{\sigma} l_L \cdot
\vec{S}_1^{\dagger}$ & $G (\overline{\nu}_{L}d_{L}^c)$ &
$\frac12 G 
(\overline{d}_{L}^c\gamma^{\mu}d_{L}^c)
(\overline{\nu}_{L}\gamma_{\mu}\nu_{L})$\\

&
$2G  (\overline{d}_{L}^c
e_{L}) (\overline{e}_{L}d_{L}^c)$ &
$G (\overline{d}_{L}^c\gamma^{\mu}d_{L}^c)
(\overline{e}_{L}\gamma_{\mu}e_{L})$\\ 

\hline

\end{tabular}
\caption{Effective four-fermion operators for scalar leptoquarks. $G$ generically stands for $\lambda^2/m_{LQ}^2$.}
\label{tab:vertices}
\end{center}
\end{table}

The LQ couplings which may contribute to $\kkbar$, $\bdbdbar$ and $\bsbsbar$ mixing should also
affect various LQ-mediated semileptonic ($b\to d(s) l^+ l^-$, $s\to d l^+l^-$) 
and purely leptonic ($B^0_{d(s)} \to l^+l^-$, $K^0\to l^+l^-$) decays. 
The estimated BRs of leptonic flavour conserving $\Delta B(S)=1$ processes 
within SM are very small compared to their experimental numbers or upper bounds, except for
$K_L\to e^+e^-, \mu^+\mu^-$. 
Therefore it is  quite reasonable to ignore the SM effects for these channels
while constraining the LQ couplings. For these mixing correlated decays, the final state leptons 
must be of the same flavour. The leptonic decay modes are theoretically clean and 
free from any hadronic uncertainties. The semileptonic modes have the usual form-factor uncertainties,
and the SM contribution cannot be neglected here.

To construct four-fermion operators from $\lambda$ type couplings which mediate leptonic and semileptonic B and K decays, one needs to integrate out the LQ field. The effective 4-fermi Hamiltonians and vertices  which are related to 
the mixing is given in Table \ref{tab:vertices}. 
The vertices show that the limits coming from leptonic or semileptonic decays will be highly correlated.
For charged leptons in the final state, one can constrain $R\tilde{S}_0$, $RS_\frac12$, $L\tilde{S}_\frac12$,
or $LS_1$ type LQs. The bounds for the first three will be the same, which is just twice that of
$LS_1$. Similarly, if we have neutrinos in the final state, $LS_0$, $L\tilde{S}_\frac12$, or $LS_1$ type
LQ couplings are bounded, all limits being the same.

The product LQ coupling may in general be complex.
If we neglect the SM, there is no scope of CP violation and the data constrains only the magnitude of
the product, so we can, if we wish, take the product to be real. In fact, if we assume CP invariance,
$K_S$ decay channels to $\ell_i^+\ell_i^-$ 
constrain only the real part of the product couplings, and $K_L$ constrains the imaginary part.
For the processes where the SM contribution cannot be neglected, we have saturated the difference
between the highest experimental prediction and the lowest SM expectation with an incoherently
summed LQ amplitude. For a quick reference, the data on the leptonic and semileptonic channels
is shown in Tables \ref{tab:Brinput2} and  \ref{tab:Brinput}. 
Note that these bounds are almost free from QCD uncertainties
except for the decay constants of the mesons, and hence are quite robust. There are other semileptonic
channels which we do not show here, e.g., $B\to K^\ast\nu\bar\nu$, because they yield less severe bounds.


\begin{table}[htbp]
\begin{center}
\begin{tabular}{||c|c||c|c||}
\hline
\hline

{ Mode } & { Branching ratio } & Mode & { Branching ratio } \\
\hline
$K_S \rightarrow e^+ e^-$ &  $< 9 \times 10^{-9}$&
$K_S \rightarrow \mu^+ \mu^-$ & $< 3.2 \times 10^{-7}$\\

$K_L \rightarrow e^+ e^-$ & $9^{+6}_{-4} \times 10^{-12}$&
$K_L \rightarrow \mu^+ \mu^-$ & $(6.84 \pm 0.11)\times 10^{-9}$\\

$B_d \rightarrow \mu^+ \mu^-$ &$<1.0 \times 10^{-8}$&
$B_d \rightarrow \mu^+ \mu^-$ &$<1.0 \times 10^{-8}$\\

$B_d \rightarrow \tau^+ \tau^-$ & $< 4.1 \times 10^{-3}$&
$B_s \rightarrow e^+ e^-$ &  $<5.4 \times 10^{-5}$\\

$B_s \rightarrow \mu^+ \mu^-$ &  $<3.3 \times 10^{-8}$&
& \\

\hline

\hline
\end{tabular}
\caption{Branching ratios for some leptonic decays  
of K and B mesons \cite{pdg}. The limits are at 90\% confidence level. The SM 
expectation is negligible.}
\label{tab:Brinput2}
\end{center}
\end{table}


\begin{table}[htbp]
\begin{center}
{\small{
\begin{tabular}{||c|c|c||c|c|c||}
\hline
\hline

{ Mode } & { Branching ratio } & SM expectation & Mode & { Branching ratio } & SM expectation\\
\hline

$K_S \rightarrow \pi^0 e^+ e^-$ & $3.0^{+1.5}_{-1.2} \times 10^{-9 }$  & $2.1\times 10^{-10}$ &
$K_S \rightarrow \pi^0 \mu^+ \mu^-$ & $2.9^{+1.5}_{-1.2} \times 10^{-9}$  & $4.8\times 10^{-10}$
\\

$K_L \rightarrow \pi^0 e^+ e^-$ &  $<2.8 \times 10^{-10}$  & $2.4\times 10^{-11}$ &
$K_L \rightarrow \pi^0 \mu^+ \mu^-$ & $<3.8 \times 10^{-10}$  & $4.4\times 10^{-12}$ \\

$K_L\rightarrow \pi^0\nu\bar\nu$ & $<6.7\times 10^{-8}$ & $(2.8\pm 0.6)\times 10^{-11}$ &
$K^+\rightarrow \pi^+\nu\bar\nu$ & $(17.3^{+11.5}_{-10.5})\times 10^{-11}$ &
$(8.5\pm 0.7)\times 10^{-11}$ \\

$K^+\rightarrow \pi^+ e^+e^-$ & $(2.88\pm 0.13)\times 10^{-7}$ &
$(2.74\pm 0.23)\times 10^{-7}$ &
$K^+\rightarrow \pi^+\mu^+\mu^-$ & $(8.1\pm 1.4)\times 10^{-8}$ &
$(6.8\pm 0.6)\times 10^{-8}$ \\

\cline{1-6}

$B_d \rightarrow \pi^0 e^+ e^-$ & $<1.4 \times 10^{-7}$  & $3.3\times 10^{-8}$ &
$B_d \rightarrow \pi^0 \mu^+ \mu^-$ &  $<1.8 \times 10^{-7}$ & $3.3\times 10^{-8}$ \\

$B_d \rightarrow K^0 e^+ e^-$ & $(1.3^{+1.6}_{-1.1}) \times 10^{-7}$ & $2.6\times 10^{-7}$  &
$B_d \rightarrow K^0 \mu^+ \mu^-$ & $(5.7^{+2.2}_{-1.8}) \times 10^{-7}$  & $(3.3\pm 0.7)\times 10^{-7}$  \\

$B_d \rightarrow K^* \mu^+ \mu^-$ & $(1.06 \pm 0.17) \times 10^{-6}$  
& $(1.0\pm 0.4)\times 10^{-6}$ &
$B_d \rightarrow K^* e^+ e^-$ & $1.39  \times 10^{-6}$  
& $(1.3\pm 0.4)\times 10^{-6}$ \\

$B_d\rightarrow \pi^0\nu\bar\nu$ & $< 2.2\times 10^{-4}$ 
& $(8.5\pm 3.5)\times 10^{-8}$ &
$B_d\rightarrow K^0\nu\bar\nu$ & $< 1.6\times 10^{-4}$ 
& $(1.35\pm 0.35)\times 10^{-5}$ \\

$B_d\rightarrow K^{*0}\nu\bar\nu$ & $< 1.2\times 10^{-4}$ 
& $3.8\times 10^{-6}$ &
$B^+\rightarrow \pi^+ e^+e^- $   &  $<8.0\times 10^{-6}$
& $(2.03\pm 0.23)\times 10^{-8}$\\

$B^+\rightarrow \pi^+ \mu^+\mu^- $   &  $<6.9\times 10^{-6}$
& $(2.03\pm 0.23)\times 10^{-8}$&
$B^+ \rightarrow \pi^+\nu\bar\nu$ & $<100  \times 10^{-6}$  
& $(9.7\pm 2.1)\times 10^{-6}$ \\

$B^+\rightarrow K^+ e^+e^- $   &  $<1.25\times 10^{-5}$
& $6.0\times 10^{-7}$ &
$B^+\rightarrow K^+ \mu^+\mu^- $   &  $<8.3\times 10^{-6}$
& $6.0\times 10^{-7}$\\

$B^+\rightarrow K^{+*} \nu\bar\nu $   &  $<8.0\times 10^{-5}$
& $(12.0\pm 4.4)\times 10^{-6}$ &
$B^+ \rightarrow K^+\nu\bar\nu$ & $<14  \times 10^{-6}$  
& $(4.5\pm 0.7)\times 10^{-6}$ \\

$B_s \rightarrow  \phi \mu^+ \mu^-$ & $(1.44\pm 0.57) \times 10^{-6}$  
& $1.6\times 10^{-6}$ &
$B_s \rightarrow  \phi \nu\bar\nu$ & $< 5.4\times 10^{-3}$  
& $(13.9\pm 5.0)\times 10^{-6}$ \\

\hline

\hline
\end{tabular}
}}
\caption{Branching ratios for some semileptonic K and B decays  
\cite{pdg,belle1,babar1,kloe,miyake}. The limits are at 90\% confidence level.
Also shown are the central values for the SM. For the SM expectations shown with an error
margin, we have taken the lowest possible values, so that the LQ bounds are most conservative.
The systematic and statistical errors have been added in quadrature.}
\label{tab:Brinput}
\end{center}
\end{table}

The constraints coming from the decay $M^0 (\equiv q_j\bar{q}_k) \to \ell_i\bar{\ell}_i$ can be expressed as
\be
|\lambda_{ij}\lambda^\ast_{ik}| < 2\sqrt{F_M} m_{LQ}^2
\ee
for $R\tilde{S}_0$, $R{S}_\frac12$, and $L\tilde{S}_\frac12$ types, and without the factor
of 2 on the righthand side for the $LS_1$ type LQs. Here
\begin{eqnarray}
F_{M} &=& \frac{1}{G_M} {\rm Br}\left(M^0 \rightarrow \ell_i\bar{\ell}_i\right),\nonumber\\
G_M &=& \frac{1}{32\pi}f_{M^0} \tau_{M^0} M_{M^0}^3 m_\ell \sqrt{1-4\frac{m_\ell^2}{M_{M^0}^2}}\,,
\end{eqnarray}
$\tau$ and $f_{M^0}$ being the lifetime and the decay constant of $M^0$ respectively.
Note that $K_L$ has a lifetime two orders of magnitude larger than that of $K_S$ and hence the bounds coming from
$K_L$ decays are going to be tighter by that amount.

\section{Numerical Inputs}
\label{sec:input}

The numerical inputs have been taken from various sources and listed in
Table \ref{tab:inputtable}. 
We use the BSW form factors \cite{F0} with a simple pole dominance, and
the relevant form factors at zero momentum transfer $q^2=0$ are taken as
follows \cite{ali}:
\bea
&&F_0^{B\to K}(0) = F_1^{B\to K}(0) = 0.38\,,\ \ 
F_0^{B\to \pi}(0) = F_1^{B\to \pi}(0) = 0.33\,,\nonumber\\
&&B\to K^\ast : V(0) = 0.37\,, A_1(0) = A_2(0) = 0.33\,, 
A_0(0) = 0.32\,,
\eea
while we take $F_0^{K\to\pi}(0) = 0.992$. This is not incompatible
with the lattice QCD result of $0.9560(84)$ \cite{lubicz}. The theoretical
uncertainty comes mostly from the form factors, but is never more than 
10\% for the LQ coupling bounds. The bounds are not a sensitive 
function of the exact values of the form factors, and remain more or
less the same even when one uses the light-cone form factors.

The mass differences $\Delta M$ are all pretty well-measured; for consistency,
we use the UTfit values \cite{utfit}. We use $\sin(2\beta_d)$ as measured in the charmonium channel \cite{hfag}.
The SM prediction is taken from the measurement of the UT sides only since
that is least likely to be affected by new physics.
(However, this need not be true always. For example, if there is a new physics
contributing in the $\bdbdbar$ 
mixing amplitude, the extracted value of $V_{td}$ may
not be equal to its SM value.)
For $\beta_s$, which is defined as ${\rm arg}(-V_{ts}V_{tb}^\ast / V_{cs}V_{cb}^\ast)$, the errors are asymmetric:
\be
\beta_s = \left( 0.47^{+0.13}_{-0.21} \right) \cup \left(1.09^{+0.21}_{-0.13}\right)\,,
\ee
which we show in a symmetrized manner. The decay constants $f_{B_{d,s}}$ are taken from \cite{ckmfitter} as
a lattice average of various groups. The same holds for $f_B\sqrt{B_B}$ and $\xi$, defined as
$\xi = f_{B_s}\sqrt{B_{B_s}} / f_{B_d}\sqrt{B_{B_d}}$, whose value we take to be
$1.258\pm 0.020 \pm 0.043$.

\begin{table}[htbp]
\begin{center}
\begin{tabular}{||c|c||c|c||}
\hline
\hline

{Observable} & {Value } & {Observable  }&{Value} \\
\hline

$\delmk $ & $5.301 \times 10^{-3}~{\rm ps}^{-1}$ & 
$|\varepsilon_K| $& $(2.228\pm 0.011) \times 10^{-3}$  \\

$\delmbd $ & $ (0.507\pm 0.005) ~{\rm ps}^{-1} $ & 
$ B_K  $ & $0.75\pm 0.07$ \\

$\delms $ & $(17.77\pm 0.12) ~{\rm  ps}^{-1}$ & 
$\eta_{B_K}$   & $1.38\pm 0.53$  \\


$\eta_{B_{B_d}(B_{B_s})}$   & $0.55\pm 0.01$ &
$ f_K $ & $160~{\rm MeV} $   \\


$\sin(2\beta_d)_{exp}$ & $0.668 \pm 0.028 $  &  
$f_{B_s}$  & $(228 \pm 17)$ MeV    \\

$\sin(2\beta_d)_{SM}$ & $0.731 \pm 0.038$ & 
$f_{B_s}/f_{B_d}$ &  $(1.199\pm 0.008 \pm 0.023)$ \\

$(\beta_s)_{exp}$ & $(0.43 \pm 0.17) \cup (1.13\pm 0.17)$ & 
$f_{B_s}\sqrt{B_{B_s}}$ & $(257 \pm 6 \pm 21)~{\rm MeV}$ \\
\hline
\hline
\end{tabular}
\caption{Input parameters. For the form factors, see text.}
\label{tab:inputtable}
\end{center}
\end{table}


\section{Analysis}

\subsection{Neutral Meson Mixing}

While our bounds are shown in Table \ref{tab:mixingbound} following
the procedure outlined in Section \ref{subsec:lqmm}, let us try to
understand the origin of these bounds.

\begin{figure*}[htbp]
\vspace{-10pt}
\centerline{\hspace{-3.3mm}
\rotatebox{-90}{\epsfxsize=6cm\epsfbox{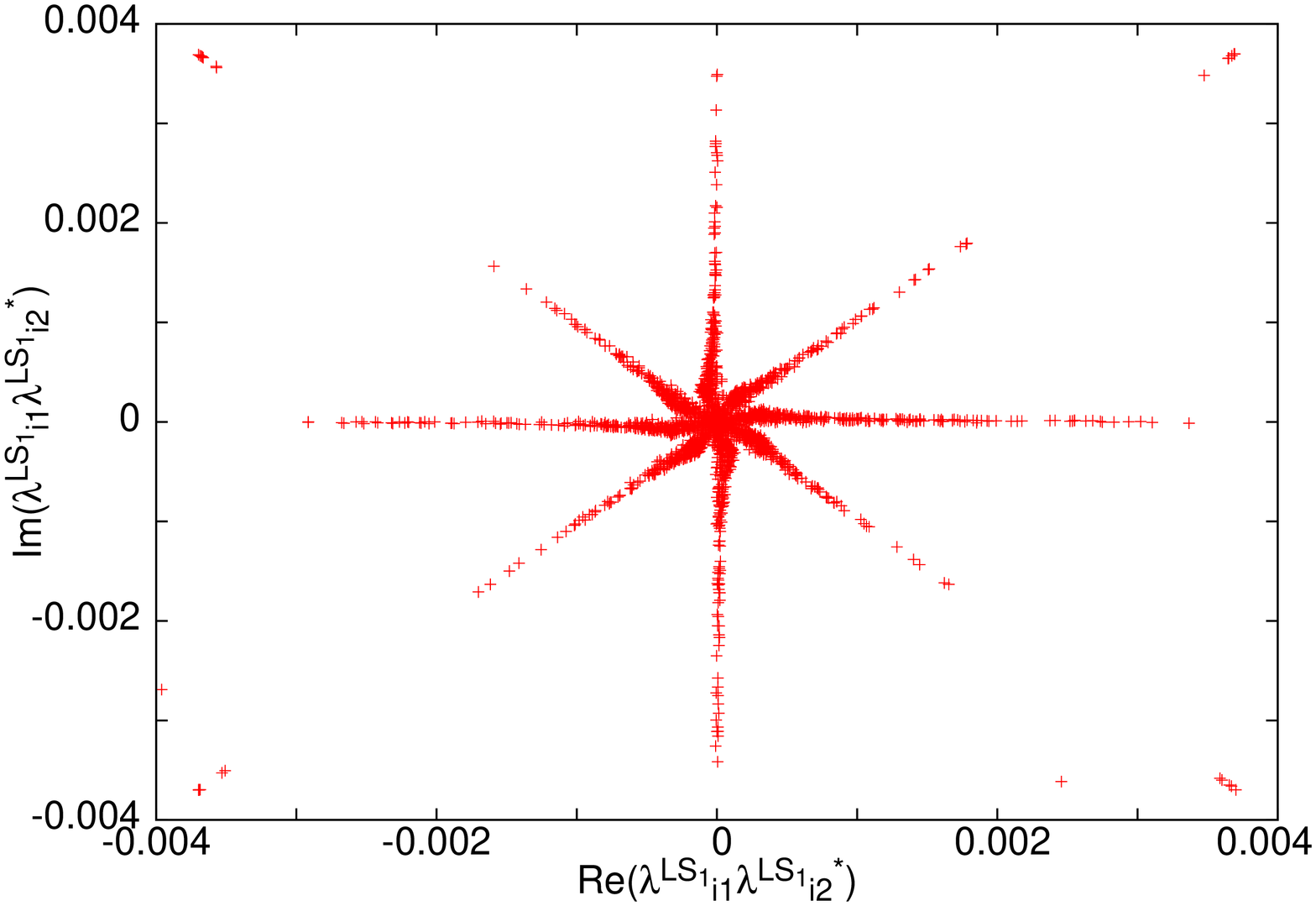}}
\hspace{-0.1cm}
\rotatebox{-90}{\epsfxsize=6cm\epsfbox{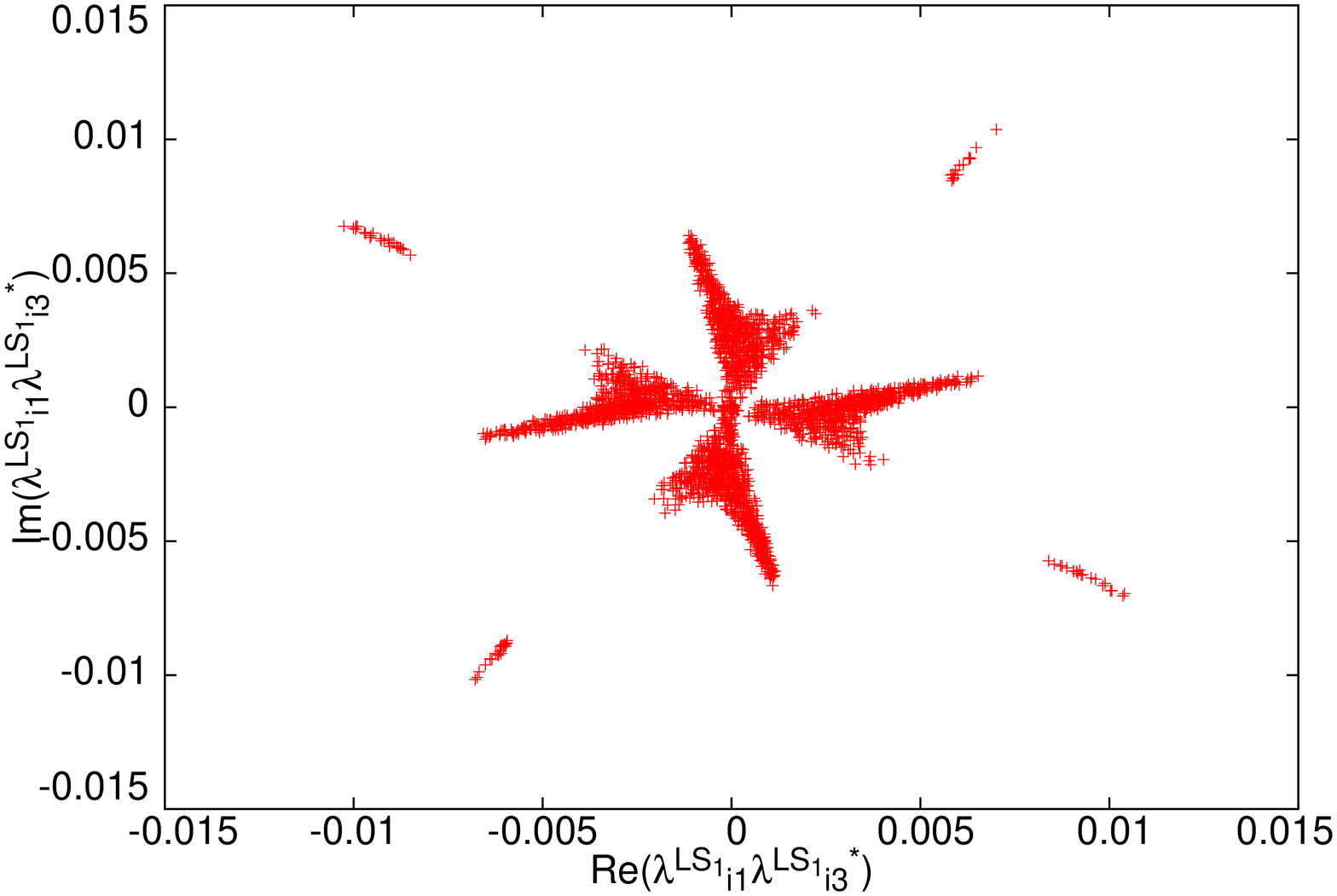}}}
\vspace*{3mm}
\centerline{\hspace{-0.5cm} (a) \hspace{7.5cm} (b)}
\hspace{3.3cm}
\caption{(a) Allowed parameter space for $\lambda_{i1}\lambda_{i2}^\ast$ for $\lambda_{LS_1}$ type
couplings. 
(b) The same for  $\lambda_{i1}\lambda_{i3}^\ast$. }
\label{fig:psmixing}
\end{figure*}

Take Figure \ref{fig:psmixing} (a) as an example, which shows the 
bounds on the real and imaginary parts of $\lambda_{i1}\lambda_{i2}^\ast$.
This is shown for the triplet LQ $S_1$; all LQs produce a similar diagram,
with the limits properly scaled. To get an idea of the scaling, one may again
look at Table \ref{tab:mixingbound}, and scale accordingly.

For the K system, we use $\Delta M_K$ and $|\varepsilon_K|$ as the
constraints. The SM part is assumed to be dominated by the short-distance
contributions only. Note the spoke-like structure; this is because 
$|\varepsilon_K|$ gives a very tight constraint on ${\rm Im}(M_{12})$ and
only those points are chosen for which $(\lambda\lambda^\ast)^2$ is
almost real. However, as we will see later, all the bounds except those for the 
$LS_0$ type LQs will be superseded
by those coming from leptonic and semileptonic K decays; however, $i=3$
bounds will stand.


A similar analysis is shown for the $\bdbdbar$ system in Figure \ref{fig:psmixing}
(b) and Table \ref{tab:mixingbound}. Note that
the bounds on the real and the imaginary parts
of any product coupling are almost the same. This is, of course, no numerical 
accident. To understand this, let us analyse the origin of these bounds. 
There are two main constraints for the $B_d$ system: $\Delta M_d$ and 
$\sin(2\beta_d)$. 
There will be a region, centred around the origin of $Re(\lambda\lambda)-Im(\lambda\lambda)$ 
plane (since $\Delta M_d$ can be explained by the SM alone),  where $|\lambda\lambda|$
is small and the phase can be arbitrary. At the $1\sigma$ level, this region appears to be
small, because the measured value of $\sin(2\beta_d)$ from the charmonium channels
is just barely compatible with that obtained from a measurement of the sides of the
unitarity triangles. The region expands if we take the error bars to be larger.
This is the SM-dominated region, 
where LQ creeps in to whatever place is left available. Any 
analysis, taking both SM and LQ but assuming incoherent sum of amplitudes,
should generate this region only.

However, there is always scope for fully constructive or destructive 
interference between SM and any NP. Consider a situation where the LQ contribution is large,
so large that even after a destructive interference with the SM amplitude,
enough is left to saturate $\Delta M_d$. This LQ-dominated region
(this is true for all NP models in general) gives us the bounds, and in the
limit where the SM can be neglected, the bounds on $Re(\lambda\lambda)$ are almost 
the same as on $Im(\lambda\lambda)$. The alignment of the fourfold symmetric
structure is different from Figure \ref{fig:psmixing} (b) because of the sizable
value of $\sin(2\beta_d)$.


\begin{table}
\begin{center} 
\begin{tabular}{||c|c|c|c|c|c||}

\hline 
      
Process & LQ Type & Real & Real part & Img part &  $\left|\lambda\lambda^*\right|$ \\
\& indices &   & Only & of Complex & of Complex & \\
\hline

$\kkbar$ & $LS_0$, $R\tilde{S}_0$, $RS_\frac12$ & 0.008 & 0.008 & 0.008 & 0.008 \\
$(i1)(i2)^\ast$ & $L\tilde{S}_\frac12$                             & 0.0055 & 0.0055 & 0.0055 & 0.0055\\
    & $LS_1$   & 0.0036  & 0.0036 & 0.0036  & 0.0036 \\
    \hline
 $\bdbdbar$ & $LS_0$, $R\tilde{S}_0$, $RS_\frac12$ & 0.009 & 0.022 & 0.022 & 0.027 \\
$(i1)(i3)^\ast$ & $L\tilde{S}_\frac12$   & 0.0063 & 0.016 & 0.016 & 0.019\\
    & $LS_1$   & 0.004  & 0.010 & 0.010 & 0.012 \\
    \hline
    $\bsbsbar$ & $LS_0$, $R\tilde{S}_0$, $RS_\frac12$ & 0.05 & 0.13 & 0.13 & 0.18 \\
$(i2)(i3)^\ast$ & $L\tilde{S}_\frac12$                             & 0.034 & 0.09 & 0.09 & 0.13\\
    & $LS_1$   & 0.02  & 0.06 & 0.06  & 0.08 \\
    \hline    
\end{tabular}
\caption{Bounds from the neutral meson mixing. The third column shows the bounds when the
couplings are assumed to be real. The last three columns are for complex couplings.}
\label{tab:mixingbound}
\end{center}
\end{table}

\begin{figure*}[htbp]
\vspace{-10pt}
\centerline{\hspace{-3.3mm}
\rotatebox{-90}{\epsfxsize=6cm\epsfbox{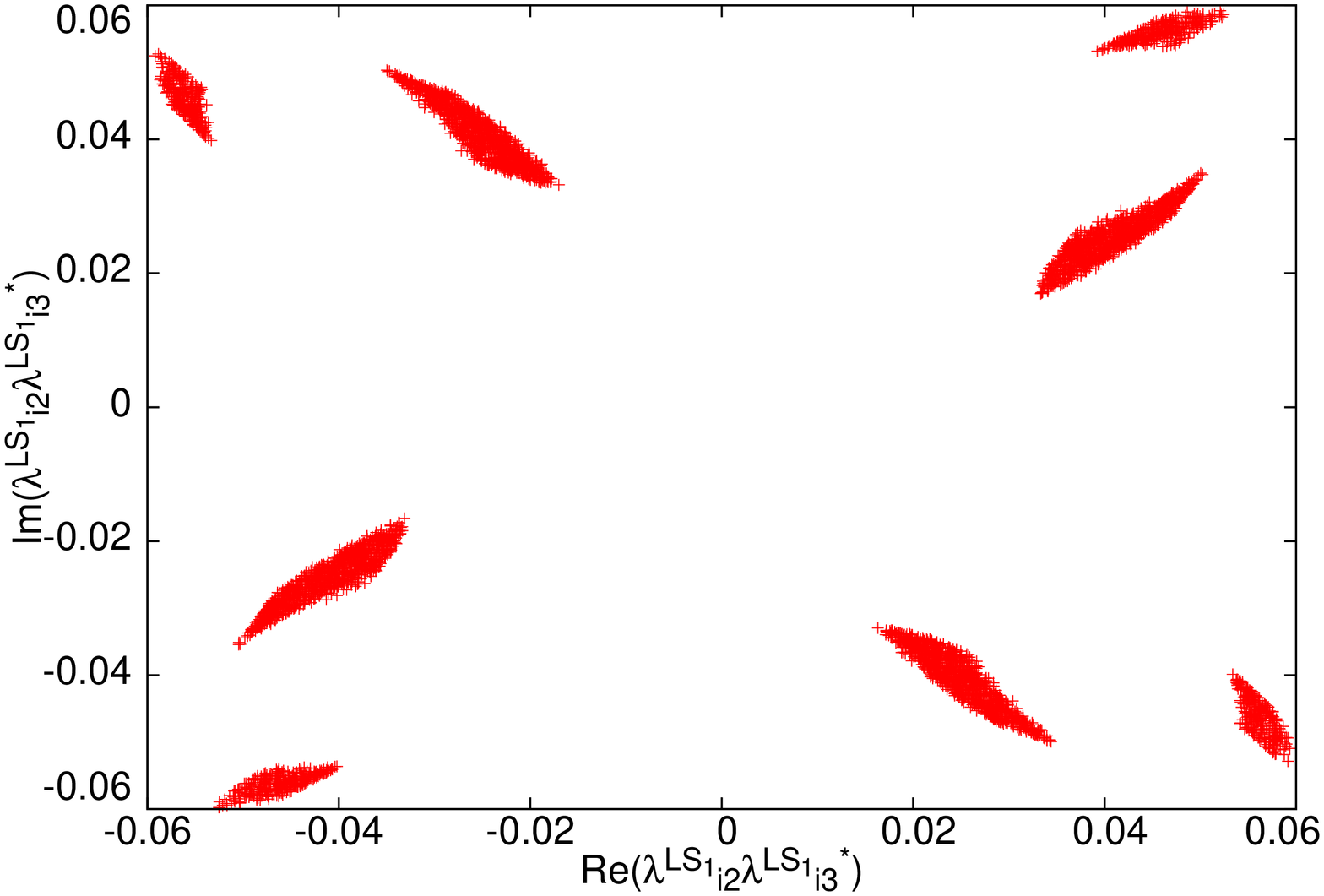}}
\hspace{-0.1cm}
\rotatebox{-90}{\epsfxsize=6cm\epsfbox{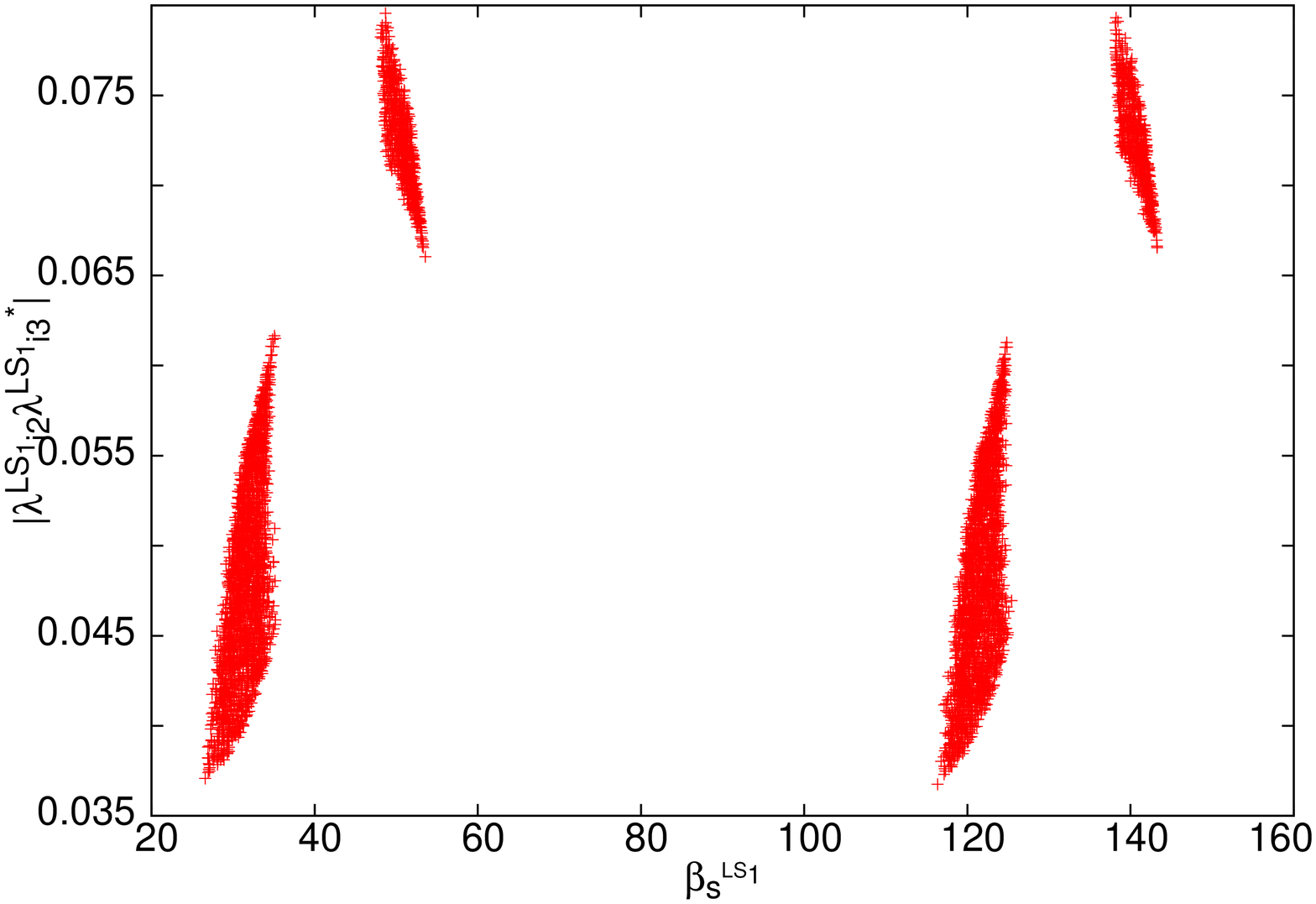}}}
\vspace*{3mm}
\centerline{\hspace{-0.5cm} (a) \hspace{7.5cm} (b)}
\hspace{3.3cm}
\caption{(a) Allowed parameter space for $\lambda_{i2}\lambda_{i3}^\ast$
(b) The reach for the angle $\beta_s$. For more details, see text.}
\label{fig:psBs}
\end{figure*}

The limits for the $B_s$ system are shown in Figure \ref{fig:psBs}. Note that
the origin is excluded at the $1\sigma$ level; this is due to the large 
observed values of $\beta_s$: $\beta_s = (25\pm 10)^\circ \cup (65\pm 10)^\circ$
in the first quadrant and a mirror image in the second quadrant. 

The magnitude of the product is bounded to be less than 0.08 at the $1\sigma$
level for the triplet LQ, and scaled according to Table \ref{tab:mixingbound}. 
For $i=2$, the relevant coupling mediates the leptonic decay $B_s\to
\mu^+\mu^-$ and semileptonic $B_d\to K^{(\ast)}\mu^+\mu^-$ decays. We will see
in the next part, just like the K system,
 that the constraints coming from such decays are much stronger.
The same observation is true for $i=1$. Again, only for $\lambda_{LS_0}$
type couplings, there is no leptonic or semileptonic contributions (the
down-type quark current couples with the neutrino current only), and the
bounds coming from the mixing stand. Thus, for $i=1,2$ and any other LQ 
except $S_0$, it is extremely improbable that the LQ contribution explains
the large mixing phase.

What happens for $i=3$? This will mediate the decays $B_s\to \tau^+\tau^-$
and $B\to X_s \tau^+\tau^-$. While there is no data on these channels yet,
we may have a consistency check with the lifetime of $B_s$. This tells us
that couplings as large as $0.05$ are allowed, but the decay $B_s\to
\tau^+\tau^-$ should be close to the discovery limit. This will be an
interesting channel to explore at the LHC. There is an exception: if 
we consider $\lambda_{LS_0}$ type couplings, neutrinos flow inside the box, and
then we have final-state neutrinos, and not $\tau$ leptons.   

Note that the box diagram with leptoquarks and leptons has a nonzero
absorptive part, which is responsible for the corresponding correlated
decays. This affects the width differences $\Delta\Gamma_{d,s}$. As has
been shown in \cite{dighe}, NP that contributes to $\Delta\Gamma$ may
enhance the mixing phase in the $\bsbsbar$ box, contrary to the Grossman
theorem \cite{grossman}, which tells that the mixing phase in the $B_s$
system must decrease due to NP if there is no absorptive amplitude in the
box diagram. The effect on $\Delta\Gamma_d/\Gamma_d$ is negligible; with
the bounds that we get here, it is never more than 1\%, or even less (note
that \cite{dighe} uses a LQ mass of 100 GeV and we need to scale their results).
For $B_s$, $\Delta\Gamma_s/\Gamma_s$ may go up to 30\% without significantly
enhancing the leptonic branching ratios like $B_s\to\tau^+\tau^-$, and one
can also get a significant nonzero phase in the $\bsbsbar$ mixing that is
indicated by the present experiments \cite{cdf-bs}.


\begin{table}
\begin{center} 
\begin{tabular}{||c|c|c||c|c|c||}

\hline
\hline 
      
{$K_{L(S)}$ Decay } &{Coupling  }& $\left|\lambda\lambda^\ast\right|$ 
&{$B_{d(s)}$  Decay } &{Coupling} &  $\left|\lambda\lambda^\ast\right|$   \\
\hline
$K_S\to e^+e^-$ & $(12)(11)^\ast$ & $1.8\times 10^{-1}$ &
$B_d\to\mu^+\mu^-$ & $(21)(23)^\ast$ & $2.8\times 10^{-3}$ \\
$K_S\to \mu^+\mu^-$ & $(22)(21)^\ast$ & $5.5\times 10^{-3}$ &
$B_d\to\tau^+\tau^-$ & $(31)(33)^\ast$ & $1.2\times 10^{-1}$ \\
$K_L\to e^+e^-$ & $(12)(11)^\ast$ & $2.4\times 10^{-4}$ &
$B_s\to\mu^+\mu^-$ & $(22)(23)^\ast$ & $4.3\times 10^{-3}$ \\
$K_L\to \mu^+\mu^-$ & $(22)(21)^\ast$ & $6.4\times 10^{-6}$ &
&&\\
\hline      
\end{tabular}

\caption{Bounds from the correlated leptonic $K_{L(S)}$ and $B_{d(s)}$  decays. The LQs are either
of $R\tilde{S}_0$, $RS_\frac12$, or $L\tilde{S}_\frac12$ type. 
For $LS_1$ type LQ, the bounds are half of that shown here.}
\label{tab:leptokbound}
\end{center}
\end{table}



\begin{table}
\begin{center} 
\begin{tabular}{||c|c|c||c|c|c||}

\hline 
\hline 
      
{Decay channel} &Coupling& $\left|\lambda\lambda^*\right|$ &
{Decay channel} &Coupling& $\left|\lambda\lambda^*\right|$\\
\hline

$K_S \rightarrow \pi^0 e^+ e^-$&$(11)(12)^*$ &$2.8 \times 10^{-3}$&
$K_L \rightarrow \pi^0 e^+ e^-$&$(11)(12)^*$ &$2.8 \times 10^{-5}$\\

$K_S \rightarrow \pi^0 \mu^+ \mu^-$&$(21)(22)^*$ &$4.6 \times 10^{-3}$&
$K_L \rightarrow \pi^0 \mu^+ \mu^-$&$(21)(22)^*$ &$5.9 \times 10^{-5}$\\

$K^+ \rightarrow \pi^+ e^+ e^-$&$(11)(12)^*$ &$1.2 \times 10^{-3}$&
$K^+ \rightarrow \pi^+ \mu^+ \mu^-$&$(21)(22)^*$ &$9.5 \times 10^{-4}$\\

$K_L \rightarrow \pi^0\nu\bar\nu$ & $(i1)(i2)^*$ & $4.6\times 10^{-4}$&
&& \\

$B_d \rightarrow \pi^0 e^+ e^-$&$(11)(13)^*$ &$1.1 \times 10^{-3}$&
$B^+ \rightarrow \pi^+ e^+ e^-$&$(11)(13)^*$ &$5.0\times 10^{-4}$\\

$B_d \rightarrow \pi^0 \mu^+ \mu^-$&$(21)(23)^*$ &$1.2 \times 10^{-3}$&
$B^+ \rightarrow \pi^+ \mu^+ \mu^-$&$(21)(23)^*$ &$4.6 \times 10^{-4}$\\

$B_d \rightarrow K^0 e^+ e^-$&$(12)(13)^*$ &$3.6 \times 10^{-4}$&
$B^+ \rightarrow K^+ e^+ e^-$&$(12)(13)^*$ &$7.0 \times 10^{-3}$\\

$B_d \rightarrow  K^* e^+e^-$&$(12)(13)^*$ &$9.7 \times 10^{-4}$&
$B_d \rightarrow  K^* \mu^+ \mu^-$&$(22)(23)^*$ &$1.1 \times 10^{-3}$\\

$B_d \rightarrow K^0 \mu^+ \mu^-$&$(22)(23)^*$ &$1.5 \times 10^{-3}$&
$B^+ \rightarrow K^+ \mu^+ \mu^-$&$(22)(23)^*$ &$5.6 \times 10^{-3}$\\

$B_d\rightarrow \pi^0\nu\bar\nu$ & $(i1)(i3)^*$ & $4.4\times 10^{-2}$&
$B^+\rightarrow \pi^+\nu\bar\nu$ & $(i1)(i3)^*$ & $2.0\times 10^{-2}$\\

$B_d\rightarrow K^0\nu\bar\nu$ & $(i2)(i3)^*$ & $2.6\times 10^{-2}$ & 
$B^+\rightarrow K^+\nu\bar\nu$ & $(i2)(i3)^*$ & $6.5\times 10^{-3}$ \\

$B_d\rightarrow K^{0*}\nu\bar\nu$ & $(i2)(i3)^*$ & $1.5\times 10^{-2}$ & 
$B^+\rightarrow K^{+*}\nu\bar\nu$ & $(i2)(i3)^*$ & $1.2\times 10^{-2}$ \\

$B_s \rightarrow \phi \mu^+ \mu^-$&$(22)(23)^*$ &$7.9 \times 10^{-4}$&
$B_s \rightarrow \phi \nu\bar\nu$&$(i2)(i3)^*$ &$9.1 \times 10^{-2}$\\

\hline

\end{tabular}
\caption{Bounds from the  correlated semileptonic B and K decays.
The LQs are either
of $R\tilde{S}_0$, $RS_\frac12$, or $L\tilde{S}_\frac12$ type. 
For $LS_1$ type LQ, the bounds are half of that shown here. For the final state neutrino
channels, the LQ can be $LS_0$, $L\tilde{S}_\frac12$, or $LS_1$ type,
all giving the same bound.}
\label{tab:semileptoBbound}
\end{center}
\end{table}

\subsection{Leptonic and Semileptonic Decays}

We have assumed only two LQ couplings to be present simultaneously, with identical lepton indices.
Thus we will be interested only in lepton flavour conserving processes. A similar analysis was done in
\cite{german} for vector LQs. 
Our bounds are shown in Table \ref{tab:leptokbound} and Table \ref{tab:semileptoBbound}. 

Apart from the leptonic $K_L$ decays, the SM amplitudes can be neglected as a first approximation.
Thus, one may saturate the experimental bounds with the LQ amplitude alone. This generates most
of the numbers in Table \ref{tab:leptokbound}. For $K_L$ decays, we consider the SM part too, and
add the amplitudes incoherently. Note that $K_L$ decays only constrain the imaginary part of
the LQ coupling. This can be understood as follows. Consider the decay $K_L\to \mu^+\mu^-$. 
While in the limit of CP invariance, one can write $K_L = (K^0-\overline{K}{}^0)/\sqrt{2}$, it is
$\lambda_{21}\lambda_{22}^\ast$ that mediates $K^0$ decay and $\lambda_{21}^\ast\lambda_{22}$ that
mediates $\overline{K}{}^0$ decay. Taking the combination, the imaginary part of the coupling
is responsible for $K_L$ decays, and the real part is responsible for $K_S$ decays.  
As mentioned earlier, $B_s\to\tau^+\tau^-$ does not have a limit yet, but the
SM expectation is about ${\cal {O}} (10^{-6})$, and if $|\lambda_{32}\lambda_{33}|
\sim 10^{-2}$, one expects the BR to be of the order of $4\times 10^{-5}$.

\begin{table} [htbp]
\begin{center}
\begin{tabular}{||c|c|c|c|c|c|c||}
\hline 
{ LQ }&{ indices}  &{ Previous }&\multicolumn {4}{|c||}{{This analysis}}\\ 
\cline{4-7} 
{ type}&&{ Bound}& \multicolumn {2}{|c|}{{ From Mixing}}&\multicolumn {2}{|c||}{{ From Decay}}\\
\cline{4-7}
&&&{Real part}&{Imag. part}& {Channel }&{ Bound}\\
\hline 
\hline
$LS_{0}$&$(i1)(i2)^*$&$1.8 \times 10^{-4}$&$ 8\times 10^{-3}$&$ 8 \times 10^{-3}$& 
$K_L\to\pi^0\nu\bar\nu$ & $\it 4.6\times 10^{-4}$\\
\cline{1-7}
$R\tilde{S_{0}}$, &$(11)(12)^*$&$2.7 \times 10^{-3}$&$8 \times 10^{-3}$&$8 \times 10^{-3}$&${K}^+ \rightarrow \pi^+ e^+ e^-$&$ \it 1.2 \times 10^{-3}$\\
&&&&&${K}_L \rightarrow \pi^0 e^+ e^-$&$\it (2.8 \times 10^{-5})$\\
\cline{2-7}
$RS_{1/2}$&$(21)(22)^*$&$5.4 \times 10^{-5}$&$8 \times 10^{-3}$&$8 \times 10^{-3}$&${K}^+ \rightarrow \pi^+ \mu^+ \mu^-$&$\it 9.5 \times 10^{-4}$\\
&&&&&${K}_L \rightarrow  \mu^+ \mu^-$&$\it 6.4 \times 10^{-6}$\\
\cline{2-7}
& $(31)(32)^*$&0.018&$ \it 8 \times 10^{-3}$&$\it 8 \times 10^{-3}$& --- & --- \\
\cline{1-7}
&$(11)(12)^*$&$1.8 \times 10^{-4}$&$5.6 \times 10^{-3}$&$5.6 \times 10^{-3}$&${K}^+ \rightarrow \pi^+ e^+ e^-$&$\it 1.2 \times 10^{-3}$\\
&&&&&${K}_L \rightarrow \pi^0 e^+ e^-$&$\it (2.8 \times 10^{-5})$\\
\cline{2-7}
$L\tilde{S}_{1/2}$&$(21)(22)^*$&$1.8 \times 10^{-4}$&$5.6 \times 10^{-3}$&$5.6 \times 10^{-3}$&${K}^+ \rightarrow \pi^+ \mu^+ \mu^-$&$\it 9.5 \times 10^{-4}$\\
&&&&&${K}_L \rightarrow  \mu^+ \mu^-$&$\it (6.4 \times 10^{-6})$\\
\cline{2-7}
&$(31)(32)^*$&$5.4 \times 10^{-5}$&$5.6 \times 10^{-3}$&$5.6 \times 10^{-3}$&
$K_L\to\pi^0\nu\bar\nu$ & $\it 4.6\times 10^{-4}$\\
\cline{1-7}
&$(11)(12)^*$&$1.8 \times 10^{-4}$&$3.6 \times 10^{-3}$&$3.6 \times 10^{-3}$&${K}^+ \rightarrow \pi^+ e^+ e^-$&$ \it 6.0 \times 10^{-4}$\\
&&&&&${K}_L \rightarrow \pi^0 e^+ e^-$&$\it (1.4 \times 10^{-5})$\\
\cline{2-7}
$LS_{1}$&$(21)(22)^*$&$2.7 \times 10^{-5}$&$3.6 \times 10^{-3}$&$3.6 \times 10^{-3}$&${K}^+ \rightarrow \pi^+ \mu^+ \mu^-$&$\it 4.8 \times 10^{-4}$\\
&&&&&${K}_L \rightarrow  \mu^+ \mu^-$&$\it (3.2 \times 10^{-6})$\\
\cline{2-7}
&$(31)(32)^*$&$1.8 \times 10^{-4}$&$ 3.6 \times 10^{-3}$&$ 3.6 \times 10^{-3}$&
$K_L\to\pi^0\nu\bar\nu$ & $\it 4.6\times 10^{-4}$\\
\cline{1-7}

\hline

\end{tabular}
\end{center}
\caption{Bounds coming from $\kkbar$ mixing and correlated decays. The better
bounds have been emphasized. Note that $K_S$ decays constrain ${\rm Re}~(\lambda_{i1}\lambda_{i2}^\ast)$
while $K^+$ decays constrain only the magnitudes; however, in view of a tight constraint on the imaginary part,
the bound from $K^+$ decay can be taken to be on the real part of the product coupling.
Here and in the next two tables, all numbers in the 
``Previous bound" column are taken from \cite{sacha}, with scaling the LQ mass to 300 GeV.} 
\label{tab:kk}
\end{table}


\begin{table} [htbp]
\begin{center}
\begin{tabular}{||c|c|c|c|c|c|c||}
\hline 
{ LQ }&{ indices}  &{ Previous }&\multicolumn {4}{|c||}{{This analysis}}\\ 
\cline{4-7} 
{ type}&&{ Bound}& \multicolumn {2}{|c|}{{ From Mixing}}&\multicolumn {2}{|c||}{{ From Decay}}\\
\cline{4-7}
&&&{Real part}&{Imag. part}& {Channel }&{ Bound}\\
\hline 
\hline
$LS_{0}$&$(i1)(i3)^*$ &0.036&{0.022}&{0.022}& 
$B^+\rightarrow\pi^+\nu\bar\nu$ & $\it 2.0\times 10^{-2}$\\
\cline{1-7}
$R\tilde{S_{0}}$, &$(11)(13)^*$&0.054&{0.022}&{0.022}&$B^+ \rightarrow \pi^+ e^+ e^-$&$\it 5.0 \times 10^{-4}$\\
\cline{2-7}
$RS_{1/2}$& $(21)(23)^*$&$7.2 \times 10^{-3}$&0.022&0.022&$B^+ \rightarrow \pi^+ \mu^+ \mu^-$&$\it 4.6 \times 10^{-4}$\\
\cline{2-7}
&$(31)(33)^*$&0.054&{\it 0.022}&{\it 0.022}&$B_d \rightarrow \tau^+ \tau^-$&$ 1.2 \times 10^{-1}$\\
\cline{1-7}
&$(11)(13)^\ast$ &0.054&0.016 & 0.016 &$B^+ \rightarrow \pi^+ e^+ e^-$&$\it 5.0 \times 10^{-4}$\\
\cline{2-7}
$L\tilde{S}_{1/2}$&$(21)(23)^\ast$&$7.2 \times 10^{-3}$&0.016&0.016&$B^+ \rightarrow \pi^+ \mu^+ \mu^-$&$\it 4.6 \times 10^{-4}$\\
\cline{2-7}
&$(31)(33)^\ast$&0.054&{\it 0.016}&{\it 0.016}&
$B^+\rightarrow\pi^+\nu\bar\nu$ & $2.0\times 10^{-2}$\\
\cline{1-7}

&$(11)(13)^*$&0.036&{0.010}&{0.010}&$B^+ \rightarrow \pi^+ e^+ e^-$&$\it 2.5 \times 10^{-4}$\\
\cline{2-7}
$LS_{1}$&$(21)(23)^*$&$3.6 \times 10^{-3}$&0.010&0.010&$B^+ \rightarrow \pi^+ \mu^+ \mu^-$&$\it 2.3 \times 10^{-4}$\\
\cline{2-7}
&$(31)(33)^*$&0.027&{0.010}&{0.010}&$B_d \rightarrow \tau^+ \tau^-$&$\it 6.2 \times 10^{-2}$\\
\cline{1-7}

\hline

\end{tabular}
\end{center}
\caption{Bounds coming from $\bdbdbar$ mixing and correlated decays. The better
bounds have been emphasized.}
\label{tab:bdbd}
\end{table}


\begin{table} [htbp]
\begin{center}
\begin{tabular}{||c|c|c|c|c|c|c||}
\hline 
{ LQ }&{ indices}  &{ Previous }&\multicolumn {4}{|c||}{{This analysis}}\\ 
\cline{4-7} 
{ type}&&{ Bound}& \multicolumn {2}{|c|}{{ From Mixing}}&\multicolumn {2}{|c||}{{ From Decay}}\\
\cline{4-7}
&&&{Real part}&{Imag. part}& {Channel }&{ Bound}\\
\hline 
\hline
$LS_{0}$&$(i2)(i3)^*$&0.36&{ 0.13}&{ 0.13}&
$B^+\rightarrow K^+\nu\bar\nu$ & $\it 6.5\times 10^{-3}$\\
\cline{1-7}
$R\tilde{S_{0}}$, &$(12)(13)^*$&$5.4 \times 10^{-3}$&0.13&0.13&$B_d \rightarrow K^0 e^+ e^-$&$\it 3.6 \times 10^{-4}$\\
\cline{2-7}
$RS_{1/2}$&$(22)(23)^*$&$7.2 \times 10^{-3}$&0.13&0.13&$B_d \rightarrow K^* \mu^+ \mu^-$&$\it 1.1\times 10^{-3}$\\
\cline{2-7}
&$(32)(33)^*$&.09&{\it 0.13}&{\it 0.13}&---&---\\
\cline{1-7}
&$(12)(13)^*$&$5.4 \times 10^{-3}$&0.09&0.09&$B_d \rightarrow K^0 e^+ e^-$&$\it 3.6 \times 10^{-4}$\\
\cline{2-7}
$L\tilde{S}_{1/2}$&$(22)(23)^*$&$7.2 \times 10^{-3}$&0.09&0.09&$B_d \rightarrow K^* \mu^+ \mu^-$&$\it 1.1\times 10^{-3}$\\
\cline{2-7}
&$(32)(33)^*$&0.054&{ 0.09}&{0.09}&
$B^+\rightarrow K^+\nu\bar\nu$ & $\it 9.3\times 10^{-3}$\\
\cline{1-7}
&$(12)(13)^*$&$2.7 \times 10^{-3}$&0.06&0.06&$B_d \rightarrow K^0 e^+ e^-$&$\it 1.8 \times 10^{-4}$\\
\cline{2-7}
$LS_{1}$&$(22)(23)^*$&$3.6 \times 10^{-3}$&0.06&0.06&$B_d \rightarrow K^* \mu^+ \mu^-$&$\it 5.5\times 10^{-4}$\\
\cline{2-7}
&$(32)(33)^*$&0.045&{0.06}&{0.06}&
$B^+\rightarrow K^+\nu\bar\nu$ & $\it 6.5\times 10^{-3}$\\
\cline{1-7}

\hline
\end{tabular}
\end{center}
\caption{Bounds coming from $\bsbsbar$ mixing and correlated decays. The better
bounds have been emphasized.}
\label{tab:bsbs}
\end{table}


For the channel $K^+\to\pi^+\nu\bar\nu$, the outgoing neutrino can have any flavour, and so
the bound is valid for $i=1,2,3$. 
However, these bounds are valid when one can have a neutrino in the final state, i.e., for
LQs of the $L$ category, which couple with lepton doublets.

Semileptonic decays give the best bounds, but they are the least robust
one, considering the uncertainty in the form factors. While we take the 
BSW form factors \cite{F0}, the lattice QCD or light-cone sum rules based
form factors may change the final results by at most 10\%. 
To be conservative, we saturate
the difference between the SM prediction and the maximum of the data
by LQ contributions.

Let us just say a few words about $B^-\to \tau^-\bar\nu$. In the SM, the
branching ratio can be worked out from eq.\ (\ref{b-lepdecay}) and is
$(9.3^{+3.4}_{-2.3})\times 10^{-5}$, where the major sources of uncertainty are
$|V_{ub}|$ and $f_B$. The observed number, $(14.3\pm 3.7)\times 10^{-5}$ 
\cite{hfag} is a bit above the SM prediction. The tension can be alleviated
with $LS_0$ or $LS_1$ type leptoquarks; the necessary combination is
$\lambda_{31}\lambda_{33}^\ast$, and the bounds that we have obtained on this particular
combination in Table \ref{tab:bdbd}
can easily jack up the branching ratio to the observed level. 
A similar exercise has been done for the leptonic $D_s$ decays in \cite{bogdan}.

We have summarized our bounds in Tables \ref{tab:kk}, \ref{tab:bdbd}, and 
\ref{tab:bsbs}. These tables contain no new information, but just shows the
best bound for a given LQ type and a given set of indices. They further show that\\
(i) Except for R-type LQs with indices (31)(33), semileptonic, and in a few cases
leptonic, decays give the best constraints. In most of the cases they are one or more
orders of magnitude stronger than those coming from the mixing, so with those
LQs, one should not expect much discernible effects from CP asymmetries.\\
(ii) While the bounds coming from decays are only on the magnitude of the product
couplings, information on the complex weak phases of these couplings must
come from mixing data, unless one makes a careful study of semileptonic CP
asymmetries.

\section{Summary and Conclusions}

In this paper we have computed the bounds on several scalar leptoquark coupling
combinations coming from $\mmbar$ mixing as well as leptonic and semileptonic
decays. 
Though such an analysis is not 
new, we have implemented several features in this analysis which were not 
been taken into account in earlier studies. 
Apart from the improved data on the B system, we have also used the
data on CP violating phases, and obtained nontrivial constraints on the
real and imaginary parts of the couplings.
We note that for the gauge-singlet LQ $S_0$ with $\lambda_{LS_0}$ type couplings,
it is possible to alleviate the mild tension between the measured and 
predicted values of $\sin(2\beta_d)$, as well as to explain the large
mixing phase in the $B_s$ system. For this type of LQs, there are no
modifications in leptonic or semileptonic channels, unless we consider 
final-state neutrinos. 

For all other type of LQs, leptonic and semileptonic decays provide the
better constraints (the exceptions are final-state $\tau$ channels). 
We do not expect any effects on nonleptonic final states like those coming
from, say, R-parity violating supersymmetry with $\lambda'$ type couplings. 
While the bounds coming from the leptonic channels are quite robust (apart from
the mild uncertainty in the meson decay constants), those coming from
semileptonic decays have an inherent uncertainty of the order of 10-15\%,
whose origin is the imprecise nature of the form factors.

\centerline{\bf{Acknowledgements}}

We thank Fabio Bossi and Jernej Kamenik for helpful comments.
The work of AK was supported by BRNS, Govt.\ of
India; CSIR, Govt.\ of India; and the DRS programme of the University Grants Commission.


\end{document}